\documentclass[twocolumn,aps,prl,preprintnumbers]{revtex4}
\usepackage{bbm}
\usepackage[latin9]{inputenc}
\usepackage{amsmath,amssymb,amsthm,amsfonts,mathrsfs}
\usepackage{graphicx}
\usepackage{txfonts}
\usepackage[colorlinks=true,citecolor=blue,linkcolor=blue,urlcolor=blue,anchorcolor=blue]{hyperref}%
\hypersetup{colorlinks=true,citecolor=blue,linkcolor=blue,urlcolor=blue}

\makeatother
\begin{document}
\title{Magnonic frequency comb through nonlinear magnon-skyrmion scattering}
\author{Zhenyu Wang$^{1}$}
\author{H. Y. Yuan$^{2}$}
\email[Corresponding author: ]{huaiyangyuan@gmail.com}
\author{Yunshan Cao$^{1}$}
\author{Z.-X. Li$^{1}$}
\author{Rembert A. Duine$^{2}$}
\author{Peng Yan$^{1}$}
\email[Corresponding author: ]{yan@uestc.edu.cn}
\affiliation{$^{1}$School of Electronic Science and Engineering and State Key Laboratory of Electronic Thin Films and Integrated Devices, University of Electronic Science and Technology of China, Chengdu 610054, China}
\affiliation{$^{2}$Institute for Theoretical Physics, Utrecht University, 3584 CC Utrecht, The Netherlands}

\begin{abstract}
An optical frequency comb consists of a set of discrete and equally spaced frequencies and has found wide applications in the synthesis over a broad range of spectral frequencies of electromagnetic waves and precise optical frequency metrology.
Despite the analogies between magnons and photons in many aspects, the analogue of an optical frequency comb in magnonic systems has not been reported. Here, we theoretically study the magnon-skyrmion interaction and find that a magnonic frequency comb (MFC) can be generated above a threshold driving amplitude, where the nonlinear scattering process involving three magnons prevails. The mode-spacing of the MFC is equal to the breathing-mode frequency of skyrmion and is thus tunable by either electric or magnetic means. The theoretical prediction is verified by micromagnetic simulations and the essential physics can be generalized to a large class of magnetic solitons. Our findings open a new pathway to observe frequency comb structures in magnonic devices that may inspire the study of fundamental nonlinear physics in spintronic platforms in the future.
\end{abstract}

\maketitle
A frequency comb is a spectrum consisting of a series of discrete and evenly spaced spectral lines.
It was originally proposed in optical systems through a pulse train generated by a mode-locked laser \cite{Udem2002} and later realized in microresonators utilizing the Kerr nonlinearity \cite{Del2007}. The optical frequency comb enables the mutual conversion between electromagnetic waves at optical and microwave frequencies and is important for high-precision frequency metrology, such as the optical atomic clock, optical ruler, exoplanets search, and molecular spectra sampling \cite{Fortier2019,Pas2018,Suh2019,Dutt2018}. The success of photonic frequency combs also inspires physicists to search for alternative frequency comb generators. Recently, phononic combs have been theoretical predicted \cite{Cao2014} and demonstrated in microscopic extensional mode resonators through a three-wave mixing process \cite{Ganesan2017}.

A magnon is the quantum of collective spin excitations in ordered magnets and it resembles the photon as a bosonic quasiparticle. Magnon spintronics rises for the long coherent time, low-power consumption and convenient electrical manipulation of magnons as information carriers. Many photonic phenomenon such as Bose-Einstein condensation, squeezed states, antibunching and multi-particle entanglement have been reported for magnonic systems \cite{Nikumi2000,Zhao2004,Bender2012,Flebus2016,Akash2019,Liu2019,yuan2020}, and cavity photon and magnon can be strongly entangled to achieve coherent information transfer \cite{Huebl2013,Gor2014,Zhang2015,Yao2017,Wolz2020,yuan202001,yuan202002,Dany2020}. Such interplay extends the horizon of traditional spintronics and further makes magnonic system a promising candidate for quantum information processing
and quantum computing. Despite these analogies, however, the analogue of frequency combs in magnonic system is yet to be realized.

\begin{figure}
  \centering
  \includegraphics[width=0.48\textwidth]{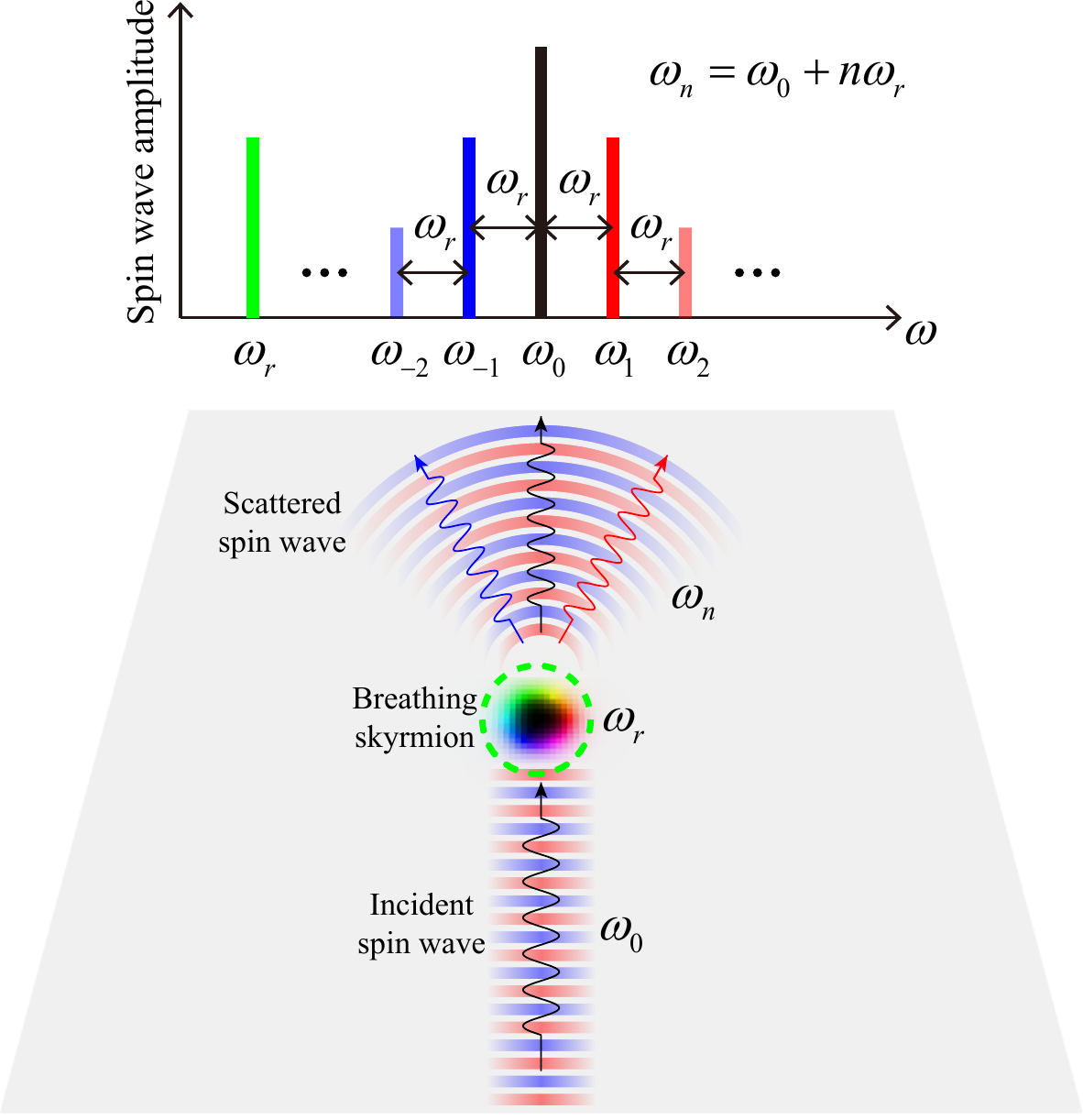}\\
  \caption{Schematic illustration of nonlinear magnon-skyrmion scattering and the resulting MFC in a magnetic film. The mode spacing in the frequncy comb is determined by the frequency of skyrmion breathing mode $\omega_r$.}\label{fig1}
\end{figure}

The leading nonlinear effects in magnetic media are three-magnon and four-magnon processes. The three-magnon process is potentially able to induce a frequency comb, but it is usually very weak due to the weak dipolar interaction. The four-magnon interaction is dominated by the strong exchange interaction, but the induced magnon spectrum is a continuous spectrum in two and higher dimensions \cite{sm}. Therefore, how to generate frequency combs remains an interesting and challenging problem.
Recently, it has been shown that magnetic textures can significantly enhance the nonlinear three-magnon interaction in magnetic systems \cite{Aristov2016,Zhang2018,Korber2020}.
Among them, magnetic skyrmions are of particular interest for both fundamental physics and potential applications in spintronic devices. Most studies on the magnon-skyrmion interactions, however, focused on the linear response, such as topological magnon Hall effect and magnon-driven skyrmion motion \cite{Mochizuki2014,Kong2013,Lin2014}. The nonlinear magnon-skyrmion interaction is rarely addressed.

In this Letter, we theoretically study the nonlinear interaction between magnetic skyrmions and magnons excited by a microwave source and predict a novel magnonic frequency comb (MFC) consisting of evenly spaced magnon modes when the amplitude of the driving field is above a threshold and its frequency lies in a frequency window. The mode-spacing in the comb is equal to the frequency of the skyrmion breathing mode. Our findings open the door for realizing the frequency comb in magnonic devices that shall extend our current understanding of the fundamental comb physics.

Let us consider a chiral magnetic system in the $xy$ plane, described by the following Hamiltonian,
\begin{equation}
\mathcal{H}=\int A(\nabla \mathbf{m})^2 - Km_z^2 +D[ m_z \nabla \cdot \mathbf{m} - (\mathbf{m} \cdot \nabla )m_z]dxdy,
\label{ham}
\end{equation}
where $\mathbf{m}$ is the normalized magnetization, $A$ is the exchange stiffness, $K$ is the easy-axis anisotropy coefficient, and $D$ is the strength of Dzyaloshinskii-Moriya
interaction (DMI). The interplay of the exchange interaction, anisotropy and DMI will stabilize the magnetic skyrmion. To investigate the interactions of magnons excited from the ground state, we make a small perturbation around the skyrmion profile, i.e.,
$\mathbf{m}(x,y)=\mathbf{m}_0(x,y) + \delta \mathbf{m}(x,y)$, where $\mathbf{m}_0(x,y)$ is the static skyrmion profile while the fluctuation $\delta \mathbf{m}(x,y)$ can be further quantized as the magnon creation and annihilation operators through Holstein-Primakoff transformation \cite{HP1940}. By substituting the ansatz back into Eq. (\ref{ham}), we
rewrite the Hamiltonian as $\mathcal{H}=\mathcal{H}^{(0)} +\mathcal{H}^{(2)}+\mathcal{H}^{(3)}+\mathcal{H}^{(4)}+...$,
where $\mathcal{H}^{(0)},\mathcal{H}^{(2)},\mathcal{H}^{(3)},\mathcal{H}^{(4)}$ are respectively the ground state energy, two, three and four magnon scattering processes \cite{sm}. Here
we observe that the main contribution to the three-magnon process is $\mathcal{H}^{(3)} \propto a_{k_1}a_{k_2}^\dagger a_{k_3}^\dagger + a_{k_1}a_{k_2} a_{k_3}^\dagger + h.c.$, where $a_{k}$ ($a_{k}^\dagger$) is the magnon annihilation (creation) operator around the skyrmion profile. Usually, this three-magnon process is weak for a dilute magnon gas excited on top of the ferromagnetic state. However, the large magnetization gradient near the skyrmion wall enhances this nonlinearity \cite{Aristov2016}. If one mode such as $a_{k_1}$ is driven with a strong power microwave, the three-magnon process may
be further amplified by the amplitude of the driving $\langle a_{k_1} \rangle $. On the other hand, a strong driving may cause significant deformation and fluctuation of the skyrmion profile, such that its internal mode including breathing and gyrotropic modes can be excited. Then the breathing mode, for instance, can hybridize with the driving mode through this three-magnon process and generate the sum and difference frequency signals, while these secondary signals can further hybridize with the breathing mode to generate higher-order frequency modes.
This chain reaction will cause a cascade of the magnonic excitation in the system, and finally result in the generation of the MFC with the frequency spacing around the skyrmion breathing mode, as illustrated in Fig. \ref{fig1}.

To capture the essential physics of the three-magnon process, we start from the Hamiltonian \cite{sm},
\begin{equation}
\begin{aligned}
\mathcal{H} &= \omega_k a_k^\dagger a_k + \omega_r a_r^\dagger a_r+ \omega_p a_p^\dagger a_p+ \omega_q a_q^\dagger a_q +g_p(a_k a_r a_p^\dagger + h.c.)\\
&+g_q (a_k a_r^\dagger a_q^\dagger + h.c.)+ h (a_k e^{i\omega_0 t} + a_k^\dagger e^{-i\omega_0 t}),
\end{aligned}
{\label{hamt}}
\end{equation}
where $\omega_k$ is the frequency of the incident spin-wave mode, $\omega_r$ is the frequency of breathing mode, $p$ and $q$ are respectively the sum-frequency mode with frequency $\omega_p=\omega_k+\omega_r$ and difference-frequency mode with frequency $\omega_q=\omega_k-\omega_r$. $g_p$ and $g_q$ are respectively the strength of the three-magnon confluence and splitting. In general, they should depend on the mode overlap of the three types of magnons. Here we treat them as phenomenological parameters. The last term is a microwave driving with amplitude $h$ and frequency $\omega_0$.

We eliminate the time dependent term in (\ref{hamt}) by transforming into a rotating frame $V= \exp [-i\omega_0 t (a_k^\dagger a_k +  a_p^\dagger a_p+  a_q^\dagger a_q)]$, and the result is,
\begin{equation}
\begin{aligned}
\mathcal{H} &= \Delta_k a_k^\dagger a_k + \omega_r a_r^\dagger a_r+ \Delta_p a_p^\dagger a_p+ \Delta_q a_q^\dagger a_q\\
&+g_p(a_k a_r a_p^\dagger + h.c.) + g_q (a_k a_r^\dagger a_q^\dagger + h.c.)
 + h (a_k + a_k^\dagger),
\end{aligned}
\label{hamwt}
\end{equation}
where the detunings are defined as $\Delta_\nu=\omega_\nu- \omega_0$ ($\nu=k,p,q$). Based on the Hamiltonian (\ref{hamwt}), we readily derive the Heisenberg dynamic equations for the system as,
\begin{subequations}
\begin{align}
&i\frac{da_k}{dt}=(\Delta_k-i\alpha_k\omega_k) a_k + g_q a_r a_q + g_p a_r^\dagger a_p + h, \\
&i\frac{da_r}{dt}=(\omega_r -i\alpha_r\omega_r)a_r + g_q a_k a_q^\dagger + g_p a_k^\dagger a_p,\\
&i\frac{da_p}{dt}=(\Delta_p -i\alpha_p\omega_p)a_p + g_p a_k a_r, \\
&i\frac{da_q}{dt}=(\Delta_q -i\alpha_q\omega_q)a_q + g_q a_k a_r^\dagger,
\end{align}\label{heisen}
\end{subequations}
where $\alpha_\nu$ ($\nu=k,r,p,q$) is the damping rate of the corresponding magnon mode that we have phenomenologically added to the equations of motion, but can be derived by assuming a Gilbert damping \cite{Gilbert2004}. In the strong driving limit, the mean field approximation can be taken such that $\langle u_iu_j \rangle =\langle u_i \rangle \langle u_j \rangle $, where $\mathbf{u} = (a_k, a_r,a_p,a_q,a_k^\dagger,a_r^\dagger,a_p^\dagger,a_q^\dagger)^T$. Then the Heisenberg equations in the steady state ($d \langle \mathbf{u} \rangle/dt=0$) always have a trivial solution:
$\langle a_k \rangle =-h/(\Delta_k-i\alpha \omega_k)$ and $\langle a_r \rangle =\langle a_p \rangle=\langle a_q \rangle=0$. This corresponds to the linear response of the system, i.e., only one magnon mode matching the driving frequency $\omega_{0}$ is strongly excited. When the driving power increases above a threshold, this trivial solution becomes unstable and the system will enter into the nonlinear regime.

To calculate the driving threshold, we expand the average of first moments around this trivial solution as
$a_\nu = \langle a_\nu \rangle + \delta a_\nu$ and rewrite the dynamic equations as $d\delta \mathbf{u}/dt=\mathbf{M}\cdot \delta \mathbf{u}$, where $\mathbf{M}$ is the drift matrix \cite{sm}. Based on the Routh-Hurwitz criteria \cite{Dejusus1987}, all eigenvalues of $\mathbf{M}$ should have negative real parts to guarantee the stability of the system. This condition implies that at least one eigenvalue of $\mathbf{M}$ is purely imaginary at the critical point, which determines the threshold field $h_{c}$ as,
\begin{equation}
h_{c}=\frac{\alpha ^2 \omega_0\sqrt{2\omega_0^3+\omega_0^2\omega_r-2\omega_0\omega_r^2-\omega_r^3}}{g\sqrt{4\omega_0 -2\omega_r}},
\end{equation}
where we assume $g_p=g_q=g$, and that all magnon modes have identical damping rates $\alpha_k=\alpha_r=\alpha_p=\alpha_q=\alpha \ll 1$ for simplicity. Here we are interested in the driving microwave with frequency much larger than the breathing-mode frequency of the skyrmion, i.e., $\omega_0 \gg \omega_r$, so that the threshold can be approximated as $h_{c} \approx \alpha^2 \omega_0^2/\sqrt{2}g$. The existence of such a threshold is the first key prediction of this letter.

Above the threshold $h_{c}$, the sum-frequency mode ($\omega_k+\omega_r$) and difference-frequency mode ($\omega_k-\omega_r$) will be excited simultaneously. By solving Eq. (\ref{heisen}), we can analytically derive the magnon population as \cite{sm},
\begin{subequations}
\begin{align}
&|\langle a_k \rangle |=\frac{h_c}{\alpha \omega_0},|\langle a_r \rangle |=\frac{\sqrt{h_c(h-h_c)}}{\alpha \omega_0},\\
&|\langle a_p \rangle |=\frac{\sqrt{h_c(h-h_c)}}{\sqrt{2}\alpha( \omega_0 + \omega_r)}, |\langle a_q \rangle |=\frac{\sqrt{h_c(h-h_c)}}{\sqrt{2}\alpha( \omega_0 - \omega_r)}.
\end{align}
\end{subequations}
We immediately see that the magnon number of the breathing mode $|\langle a_r \rangle |^2$ will keep accumulating as we increase the driving field and even surpass the driving magnons. In reality, this trend will be suppressed due to the four-magnon scattering ($\mathcal{H}^{(4)}$ term). Within mean-field formalism, the four magnon interaction will modify the eigenfrequency $\omega_r$ as $\omega_r + \kappa\langle a_r^\dagger a_r \rangle$, where the coefficient $\kappa$ represents the interaction strength. Now the steady amplitude is approximately $|\langle a_k \rangle | \sim h/(\alpha \omega_k), |\langle a_r \rangle | \sim \epsilon \omega_r / \kappa$, where the dimensionless parameter $\kappa$ can be determined by solving a set of self-consistent equations \cite{sm}.
Below, we shall verify these theoretical predictions by numerically simulating the interaction of skyrmions with magnons using $\mathrm{Mumax}^3$ \cite{mumax}.

\begin{figure}
  \centering
  \includegraphics[width=0.5\textwidth]{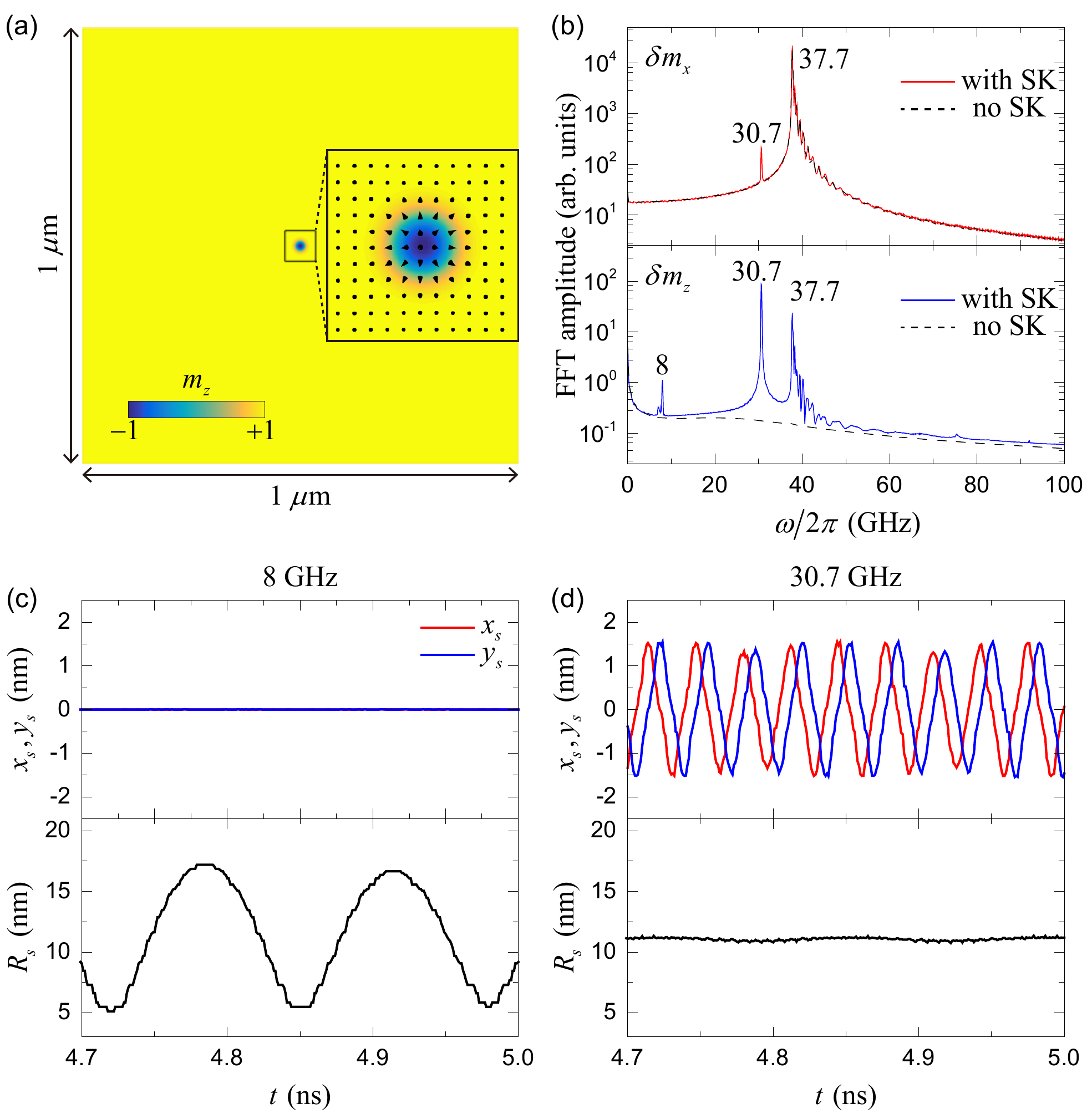}\\
  \caption{(a) Schematic illustration of a magnetic thin film hosting a N\'{e}el-type skyrmion. (b) The internal spectrum of the magnetic film with (solid line) and without skyrmion (dashed line). (c)(d) The time evolution of skyrmion position and skyrmion radius under the driving with frequency $\omega/2\pi =8$ and $30.7$ GHz, respectively.}\label{fig2}
\end{figure}

We consider a ferromagnetic thin film shown in Fig. \ref{fig2}(a) with dimensions $1000 \times 1000 \times 1~\mathrm{nm}^3$. It can host a N\'{e}el-type skyrmion at the film center, if the interfacial symmetry is broken by growing a heavy metal layer in adjacent to the magnetic film. The magnetization dynamics is simulated by numerically solving the Landau-Lifshitz-Gilbert equation,
\begin{equation}
\frac{\partial \mathbf{m}}{\partial t}=-\gamma \mathbf{m} \times \mathbf{h}_{\mathrm{eff}} + \alpha \mathbf{m} \times \frac{\partial \mathbf{m}}{\partial t},
\end{equation}
where $\gamma$ is gyromagnetic ratio, $\mathbf{h}_{\mathrm{eff}}=-\delta \mathcal{H}/\delta \mathbf{m}$ is the effective field including exchange, anisotropy, dipolar, and driving AC fields, and $\alpha$ is Gilbert damping. The magnetic parameters of Co \cite{Co} are used in the simulations \cite{sm}. The absorbing boundary condition is adopted to minimize the spin-wave reflection by film edges \cite{Venkat2018}.

We first characterize the intrinsic magnon excitation spectrum of the skyrmion by applying a sinc-function field in the film plane,
i.e., $\mathbf{h}(t)= h_0\mathrm{sinc}(\omega_{H} t)\hat{x} $ with amplitude $h_0 = 10$ mT and cut-off frequency $\omega_{H}/2\pi =100$ GHz. The internal spectrum of the magnetic system is obtained by carrying a standard fast Fourier transform (FFT) for each cell and then averaging over the whole film. Three main peaks centering at 37.7 GHz, 30.7 GHz, and 8 GHz are found in the $\delta m_{z}$ spectrum, as shown in Fig. \ref{fig2}(b), while, in the absence of skyrmion, only the 37.7 GHz mode appears in the $\delta m_{x}$ spectrum. Through this comparison, 37.7 GHz mode is ascertained to be the ferromagnetic resonance (FMR) mode, whose frequency can be analytically calculated as $\omega_c/2\pi=\gamma(2K-\mu_0M_s^2)/(2\pi M_s) = 37.5$ GHz. We notice a foldover of FMR lineshape to lower frequency under the influence of canted magnetization at the edge. Moreover, by analyzing the skyrmion motion under the sinusoidal microwave driving of 8 GHz and 30.7 GHz, as shown in Fig. \ref{fig2}(c) and \ref{fig2}(d), we identify them as the breathing mode and counterclockwise gyrotropic mode, respectively.

\begin{figure}
  \centering
  \includegraphics[width=0.5\textwidth]{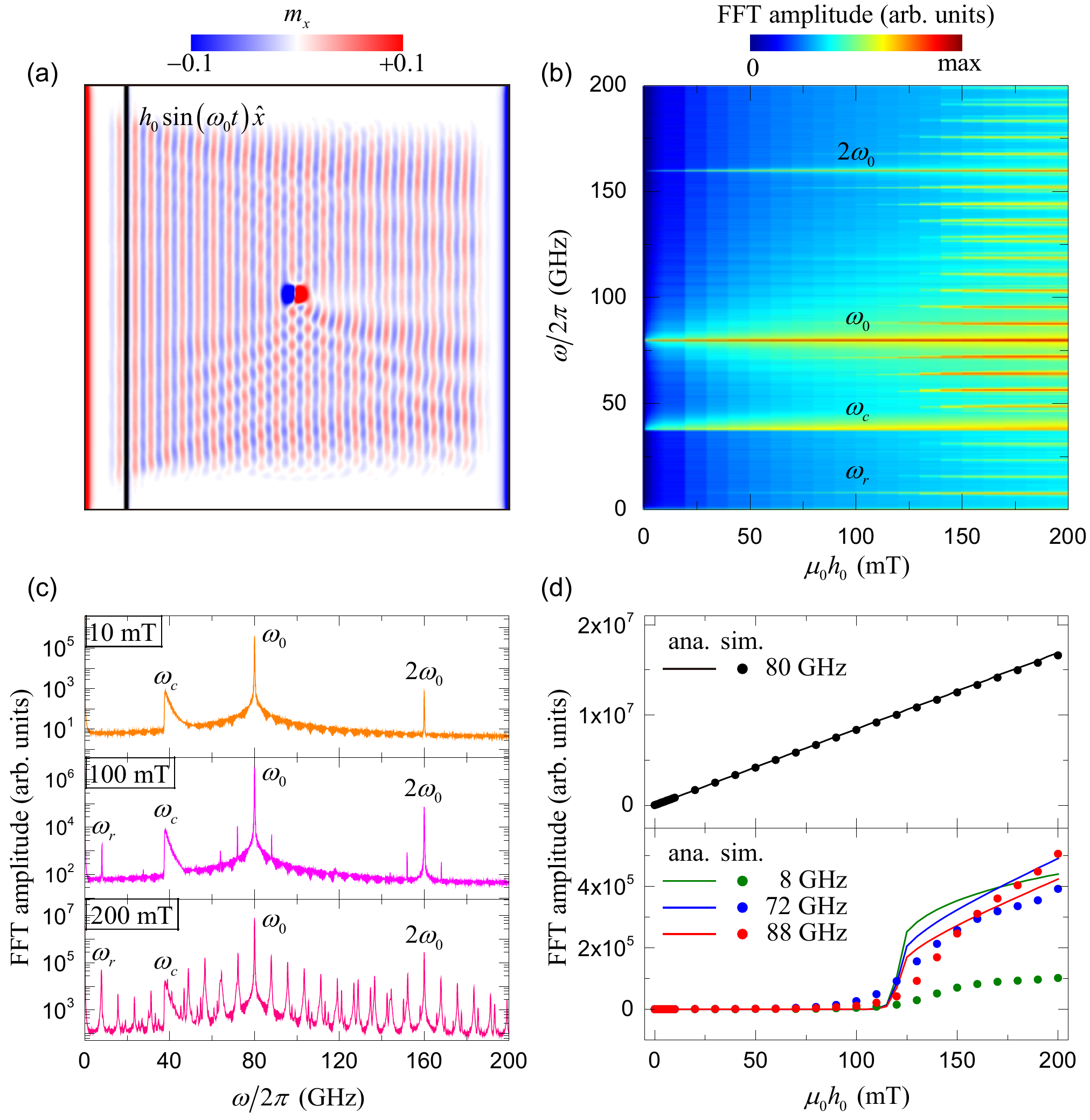}\\
  \caption{(a) Illustration of magnetic film subjected to a microwave source applied in a rectangular region (black bar). (b) The complete response of the system as we tune the driving amplitude ($h_0$). The driving frequency is fixed at 80 GHz. The color codes the amplitude of the excitation. (c) Response of the system at three representative fields $\mu_0h_0=10$ mT (upper panel), 100 mT (middle panel) and 200 mT (lower panel), respectively. (d) The amplitude of the four main peaks at $\omega/2\pi=80, ~8, $ and $80 \pm 8$ GHz as a function of driving amplitude obtained by integrating in 0.5 GHz wide frequency window.  Analytical curves are calculated by solving the Heisenberg equation with parameters: $\alpha=0.001, \omega_k/2\pi=80~\mathrm{GHz}, \omega_r/2\pi=8 ~\mathrm{GHz},$ and $g= 4.5\times10^{-3} ~\mathrm{GHz}$.}\label{fig3}
\end{figure}

According to the theoretical rationale, to generate a MFC, one needs a strong enough driving. Here we apply a microwave driving field $\mathbf{h}(t)= h_0\mathrm{sin}(\omega_0 t)\hat{x}$ with $\omega_0/2\pi=80$ GHz in a narrow rectangular area of the film [black bar in Fig. \ref{fig3}(a)]. The magnons that are excited near the source will propagate toward the skyrmion located at the film center and interact with it. To analyze the magnon excitation in the skyrmion region, we make FFT of the time dependent magnetization and record the response at various fields. Figure \ref{fig3}(b) shows the response distribution in frequency space as we tune the driving amplitude $h_0$, and three phases can be classified: (i) Below 100 mT, mainly the magnon mode matching the driving frequency $\omega_0/2\pi = 80$ GHz and the FMR mode ($\omega_c$) are excited. (ii) In the window from 100 to 130 mT, two main side-peaks at $80+8=88$ GHz and $80-8=72$ GHz are excited. (iii) Above 130 mT, clear frequency comb with mode-spacing of the breathing-mode frequency emerges, as shown in Fig. \ref{fig3}(c). These features are consistent with the theoretical predictions. It is noted that multiple frequency lines between MFC centered at 80 GHz and the double frequency mode 160 GHz overlap each other, which is due to the slight deviation of breathing mode from 8 GHz under strong driving. The mode splitting-like behavior near 200 GHz (Nyquist frequency) is a false effect of the FFT analysis \cite{sm}.
Figure \ref{fig3}(d) shows the quantitative comparison between theory and simulations for the amplitudes of four main modes at 80 GHz (black dots), 8 GHz (green dots), 72 GHz (blue dots), and 88 GHz (red dots), respectively. The analytical model thus captures the trend of these four modes well, while the main deviation is on the amplitude of breathing mode. This is attributed to the simplified three-mode interaction in the theoretical formalism, and the breathing mode may be further consumed by generating the higher-order modes inside the MFC.

\begin{figure}
  \centering
  \includegraphics[width=0.5\textwidth]{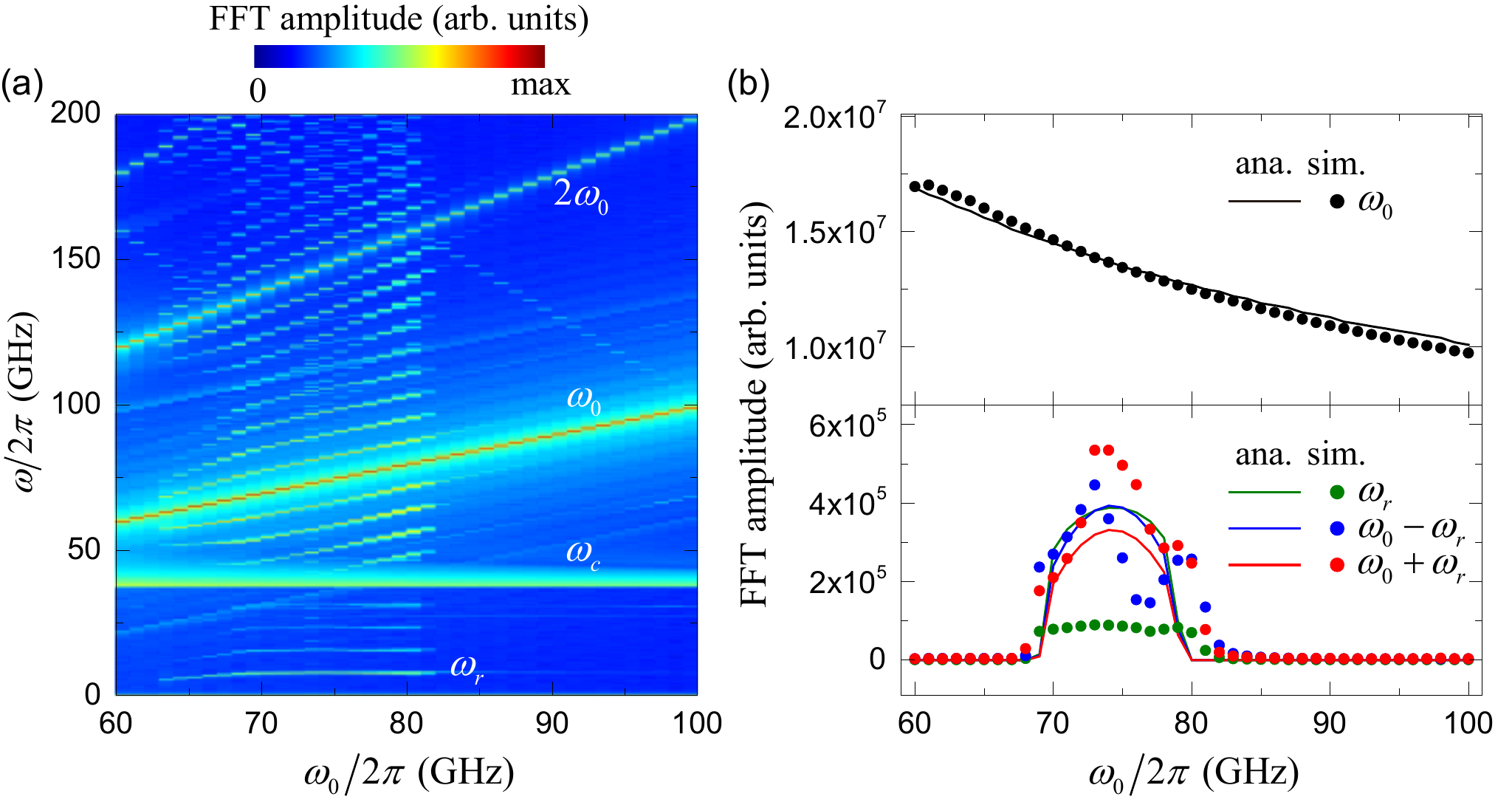}\\
  \caption{(a) Response of the system as a function of the frequency of the microwave source. The driving amplitude is fixed at 150 mT. (b) The amplitudes of the four main modes $\omega_0, \omega_0 \pm \omega_r, \omega_r$ as a function of driving amplitude.  The theoretical curves are calculated resembling that in Fig. \ref{fig3} with a driving frequency dependent coupling $g$. $g_0=6.65\times10^{-3} ~\mathrm{GHz}, \sigma = 8~\mathrm{GHz}$. }\label{fig4}
\end{figure}

To study the robustness of the magnonic comb against fluctuations of the driving frequency, we vary the microwave frequency $\omega_0/2\pi$ ranging from 60 up to 100 GHz. Figure \ref{fig4}(a) shows that the magnonic comb is mostly visible in a window from 67 to 82 GHz, while the mode spacing of the comb is always equal to the frequency of breathing mode (8 GHz) regardless of the driving frequency. To understand this behavior, we notice that the window is around the double frequency of the FMR mode ($2\omega_c/2\pi$=75 GHz). This suggests that the spins near the skyrmion may be more effectively excited though subharmonic process. Such consideration naturally implies that the strength of three magnon-process ($g_p,g_q$) will be frequency dependent. Here we assume that it follows a Gaussian profile as $g_p(\omega)=g_q(\omega)=g_0 e^{-(\omega -2\omega_c)^2/\sigma^2}$ with $g_0$ being determined by the field threshold shown in Fig. \ref{fig3}(d). Figure \ref{fig4}(b) shows that this consideration can recover the excitation window of frequency comb very well. Meanwhile, the amplitude of four main modes modes $\omega_0$ (black dots), $\omega_r$ (green dots), $\omega_0-\omega_r$ (blue dots), $\omega_0+\omega_r$ (red dots) as a function of the driving frequency can be described quantitatively. The breathing mode $\omega_r$ (green lines) is higher than numerical values, which, similar to Fig. \ref{fig3}, is due to the neglect of higher-order modes in our theoretical formalism. As a comparison, the frequency comb is absent in a ferromagnetic state under microwave driving \cite{sm}.

To verify our theoretical predictions in experiments, one can locally inject a spin wave in a chiral magnetic thin film hosting skyrmions, and then measure the spectrum of the excited spin waves near the skyrmion through frequency-resolved techniques such as Brillouin light scattering \cite{Seb2015}. This allows the real-time identification of the frequency modes in the magnonic comb. Moreover, since the breathing-mode frequency of a skyrmion is inversely proportional to the square of skyrmion size ($R_s$), i.e., $\omega_r \propto R_s^{-2}$ \cite{Kra2018}, we estimate that the mode spacing in a magnonic comb can range from GHz regime for a 10-nm skyrmion down to kilohertz regime for a 1000-nm skyrmion. This complements the working frequency of the optical frequency comb with the mode-spacing from 50 MHz to 500 GHz \cite{Fortier2019}, and can be potentially extended to the frequency range of THz in antiferromagnets. The MFC technology can be utilized in spin wave excitation and calibration \cite{Muralidhar2021}, accurate detection of magnetic solitons and defects by analyzing the comb spectroscopy before and after passing through the magnetic media, and even find its role in engineering phase-coherent magnon laser \cite{Zhang2018,Liu2020} and magnon quantum computing \cite{Andrianov2014,Francis2020}. The present study uncovers the important physics associated with the internal excitations of skyrmions, which should shed a new light on skyrmionics.

In conclusion, we have shown that a strong microwave driving can induce a hybridization of the driving mode with the breathing mode of a magnetic skyrmion, generating a magnonic frequency comb. The threshold of the driving amplitude and the excitation window of the driving frequency are well modelled by our Hamiltonian formalism involving the sum-frequency and difference-frequency processes of three magnons. We envision that the essential physics should also hold for other magnetic solitons with low-frequency internal modes, such as antiskyrmions, domain walls, and vortices. Note that the experimental study of the interaction of spin-wave and domain wall was recently reported \cite{Han2020}. Terahertz frequency combs induced by nonlinear magnon-solition interaction in antiferromagnets are also an interesting issue. Our results open a novel avenue to study the frequency comb physics in magnetic systems combining the advantages of magnons and general magnetic textures.

\begin{acknowledgments}
This work was funded by the National Natural Science Foundation of China (Grants No. 12074057, No. 11604041, No. 11704060, and No. 11904048). Z.W. and Z.X.L. acknowledge the financial support from the China Postdoctoral Science Foundation (Grant No. 2019M653063 and No. 2019M663461). R.A.D. has received funding from the European Research Council (ERC) under the European Unions Horizon 2020 research and innovation programme (Grant No. 725509).

{\it Note added.}---Recently, we became aware of a preprint \cite{Hula2021} which experimentally demonstrated the generation of spin-wave frequency combs.

Z.W. and H.Y.Y. contributed equally to this work.
\end{acknowledgments}


\begin{thebibliography}{99}
\bibitem{Udem2002} Th. Udem, R. Holzwarth, and T. W. H\"{a}nsch, Optical frequency metrology, \href{https://doi.org/10.1038/416233a}{Nature \textbf{416}, 233 (2002)}.

\bibitem{Del2007} P. Del'Haye, A. Schliesser, O. Arcizet, T. Wilken, R. Holzwarth, and T. J. Kippenberg, Optical frequency comb generation from a monolithic micoresonator, \href{https://doi.org/10.1038/nature06401}{Nature \textbf{450}, 1214 (2007)}.

\bibitem{Fortier2019} T. Fortier and E. Baumann, 20 years of developments in optical frequency comb technology and applications, \href{https://doi.org/10.1038/s42005-019-0249-y}{Commun. Phys. \textbf{2}, 153 (2019)}.

\bibitem{Pas2018} A. Pasquazi, M. Peccianti, L. Razzari, D. J. Moss, S. Coen, M. Erkintalo, Y. K. Chembo, T. Hansson, S. Wabnitz, P. De$\mathrm{l}^{'}$Haye, X. Xue, A. M. Weiner, and R. Morandotti, Micro-combs: A novel generation of optical sources, \href{https://doi.org/10.1016/j.physrep.2017.08.004}{Phys. Rep. \textbf{729}, 1 (2018)}.

\bibitem{Suh2019} M. G. Suh, X. Yi, Y. H. Lai, S. Leifer, I. S. Grudinin, G. Vasisht, E. C. Martin, M. P. Fitzgerald, G. Doppmann, J. Wang, D. Mawet, S. B. Papp, S. A. Diddams, C. Beichman, and K. Vahala, Searching for exoplanets using a microresonator astrocomb, \href{https://doi.org/10.1038/s41566-018-0312-3}{Nat. Photon. \textbf{13}, 25 (2019)}.

\bibitem{Dutt2018} A. Dutt, C. Joshi, X. Ji, J. Cardenas, Y. Okawachi, K. Luke, A. L. Gaeta, and M. Lipson, On-chip dual-comb source for spectroscopy, \href{https://doi.org/10.1126/sciadv.1701858}{Sci. Adv. \textbf{4}, e1701858 (2018)}.

\bibitem{Cao2014} L. S. Cao, D. X. Qi, R. W. Peng, M. Wang, and P. Schmelcher, Phononic Frequency comb through nonlinear resonance, \href{https://doi.org/10.1103/PhysRevLett.112.075505}{Phys. Rev. Lett. \textbf{112}, 075505 (2014)}.

\bibitem{Ganesan2017} A. Ganesan, C. Do, and A. Seshia, Phononic Frequency comb via intrinsic three-wave mixing, \href{https://doi.org/10.1103/PhysRevLett.118.033903}{Phys. Rev. Lett. \textbf{118}, 033903 (2017)}.

\bibitem{Nikumi2000} T. Nikumi, M. Oshikawa, A. Oosawa, and H. Tanaka, Bose-Einstein Condensation of Dilute Magnons in $\mathrm{TlCuCl}_{3}$, \href{https://doi.org/10.1103/PhysRevLett.84.5868}{Phys. Rev. Lett. \textbf{84}, 5868 (2000)}.

\bibitem{Zhao2004} J. Zhao, A. V. Bragas, D. J. Lockwood, and R. Merlin, Magnon Squeezing in an Antiferromagnet: Reducing the Spin Noise below the Standard Quantum Limit, \href{https://doi.org/10.1103/PhysRevLett.93.107203}{Phys. Rev. Lett. \textbf{93}, 107203 (2004)}.

\bibitem{Bender2012} S. A. Bender, R. A. Duine, and Y. Tserkovnyak, Electronic Pumping of Quasiequilibrium Bose-Einstein-Condensed Magnons, \href{https://doi.org/10.1103/PhysRevLett.108.246601}{Phys. Rev. Lett. \textbf{108}, 246601 (2012)}.

\bibitem{Flebus2016} B. Flebus, S. A. Bender, Y. Tserkovnyak, and R. A. Duine, Two-Fluid Theory for Spin Superfluidity in Magnetic Insulators, \href{https://doi.org/10.1103/PhysRevLett.116.117201}{Phys. Rev. Lett. \textbf{116}, 117201 (2016)}.

\bibitem{Akash2019} A. Kamra, E. Thingstad, G. Rastelli, R. A. Duine, A. Brataas, W. Belzig, and A. Sudb{\o}, Antiferromagnetic magnons as highly squeezed Fock states underlying quantum correlations, \href{https://doi.org/10.1103/PhysRevB.100.174407}{Phys. Rev. B \textbf{100}, 174407 (2019)}.
\bibitem{Liu2019} Z.-X. Liu, H. Xiong, and Y. Wu, Magnon blockade in a hybrid ferromagnet-superconductor quantum system, \href{https://doi.org/10.1103/PhysRevB.100.134421}{Phys. Rev. B \textbf{100}, 134421 (2019)}.

\bibitem{yuan2020} H. Y. Yuan and R. A. Duine, Magnon antibunching in a nanomagnet, \href{https://doi.org/10.1103/PhysRevB.102.100402}{Phys. Rev. B \textbf{102}, 100402(R) (2020)}.

\bibitem{Huebl2013} H. Huebl, C. W. Zollitsch, J. Lotze, F. Hocke, M. Greifenstein, A. Marx, R. Gross, and S. T. B. Goennenwein, High Cooperativity in Coupled Microwave Resonator Ferrimagnetic Insulator Hybrids, \href{https://doi.org/10.1103/PhysRevLett.111.127003}{Phys. Rev. Lett. \textbf{111}, 127003 (2013)}.

\bibitem{Gor2014} M. Goryachev, W.G. Farr, D. L. Creedon, Y. Fan, M. Kostylev, and M.E. Tobar, High-Cooperativity Cavity QED with Magnons at Microwave Frequencies, \href{https://doi.org/10.1103/PhysRevApplied.2.054002}{Phys. Rev. Applied \textbf{2}, 054002 (2014)}.

\bibitem{Zhang2015} X. Zhang, C.-L. Zou, N. Zhu, F. Marquardt, L. Jiang, and H. X. Tang, Magnon dark modes and gradient memory, \href{https://doi.org/10.1038/ncomms9914}{Nat. Commun. \textbf{6}, 8914 (2015)}.

\bibitem{Yao2017} B. Yao, Y. S. Gui, J. W. Rao, S. Kaur, X. S. Chen, W. Lu, Y. Xiao, H. Guo, K.-P. Marzlin, and C.-M. Hu, Cooperative polariton dynamics in feedback-coupled cavities, \href{https://doi.org/10.1038/s41467-017-01796-7}{Nat. Commun. \textbf{8}, 1437 (2017)}.

\bibitem{Wolz2020} T. Wolz, A. Stehli, A. Schneider, I. Boventer, R. Mac\^{e}do, A. V. Ustinov, M. Kl\"{a}ui, and M. Weides, Introducing coherent time control to cavity magnon-polariton modes, \href{https://doi.org/10.1038/s42005-019-0266-x}{Commun. Phys. \textbf{3}, 3 (2020)}.

\bibitem{Dany2020} D. Lachance-Quirion, S. P. Wolski, Y. Tabuchi, S. Kono, K. Usami, and Y. Nakamura, Entanglement-based single-shot detection of a single magnon with a superconducting qubit, \href{https://doi.org/10.1126/science.aaz9236}{Science \textbf{367}, 425 (2020)}.

\bibitem{yuan202001} H. Y. Yuan, Shasha Zheng, Z. Ficek, Q. Y. He, and M.-H. Yung, Enhancement of magnon-magnon entanglement inside a cavity, \href{https://doi.org/10.1103/PhysRevB.101.014419}{Phys. Rev. B \textbf{101}, 014419 (2020)}.

\bibitem{yuan202002} H. Y. Yuan, P. Yan, S. Zheng, Q. Y. He, K. Xia, and M.-H. Yung, Steady Bell State Generation via Magnon-Photon Coupling, \href{https://doi.org/10.1103/PhysRevLett.124.053602}{Phys. Rev. Lett. \textbf{124}, 053602 (2020)}.

\bibitem{sm} See Supplemental Material at http://link.aps.org/supplemental/ for an analytical derivation of the three-magnon interaction, the explicit form of drift matrix and the resulting magnon distribution, magnon spectrum induced by four-magnon process, materials parameters in micromagnetic simulations, the generation of magnonic frequency comb under microwave driving with higher frequencies, the origin of the higher-order modes in the comb, absence of frequency comb in a uniform ferromagnetic state and spin wave profiles and skyrmion motion in the MFC, which includes Refs. \cite{Aristov2016,Zhang2018,HP1940,Dejusus1987}.

\bibitem{Aristov2016} D. N. Aristov and P. G. Matveeva, Stability of a skyrmion and interaction of magnons, \href{https://doi.org/10.1103/PhysRevB.94.214425}{Phys. Rev. B \textbf{94}, 214425 (2016)}.

\bibitem{Zhang2018} B. Zhang, Z. Wang, Y. Cao, P. Yan, and X. R. Wang, Eavesdropping on spin waves inside the domain-wall nanochannel via three-magnon processes, \href{https://doi.org/10.1103/PhysRevB.97.094421}{Phys. Rev. B \textbf{97}, 094421 (2018)}.

\bibitem{Korber2020} L. K\"{o}ber, K. Schultheiss, T. Hula, R. Verba, J. Fassbender, A. K\'{a}kay, and H. Schultheiss, Nonlocal Stimulation of Three-Magnon Splitting in a Magnetic Vortex, \href{https://doi.org/10.1103/PhysRevLett.125.207203}{Phys. Rev. Lett. \textbf{125}, 207203 (2020)}.
%
\bibitem{Mochizuki2014} M. Mochizuki, X. Z. Yu, S. Seki, N. Kanazawa, N. Kanazawa, J. Zang, M. Mostovoy, Y. Tokura, and N. Nagaosa, Thermally driven ratchet motion of a skyrmion microcrystal and topological magnon Hall effect, \href{https://doi.org/10.1038/nmat3862}{Nat. Mater. \textbf{13}, 241 (2014)}.
\bibitem{Kong2013} L. Kong and J. Zang, Dynamics of an Insulating Skyrmion under a Temperature Gradient, \href{https://doi.org/10.1103/PhysRevLett.111.067203}{Phys. Rev. Lett. \textbf{111}, 067203 (2013)}.
\bibitem{Lin2014} S.-Z. Lin, C. D. Batista, C. Reichhardt, and A. Saxena, ac Current Generation in Chiral Magnetic Insulators and Skyrmion Motion induced by the Spin Seebeck Effect, \href{https://doi.org/10.1103/PhysRevLett.112.187203}{Phys. Rev. Lett. \textbf{112}, 187203 (2014)}.

\bibitem{HP1940} T. Holstein and H. Primakoff, Field Dependence of the Intrinsic Domain Magnetization of a Ferromagnet, \href{https://doi.org/10.1103/PhysRev.58.1098}{Phys. Rev. \textbf{58}, 1098 (1940)}.

\bibitem{Gilbert2004} T. L. Gilbert, A phenomenological theory of damping in ferromagnetic materials, \href{https://doi.org/10.1109/TMAG.2004.836740}{IEEE Trans. Magn. \textbf{40}, 3443 (2004)}.

\bibitem{Dejusus1987} E. X. DeJesus and C. Kaufman, Routh-Hurwitz criterion in the examination of eigenvalues of a system of nonlinear ordinary differential equations, \href{https://doi.org/10.1103/PhysRevA.35.5288}{Phys. Rev. A \textbf{35}, 5288 (1987)}.


\bibitem{mumax} A. Vansteenkiste, J. Leliaert, M. Dvornik, M. Helsen, F. Garcia-Sanchez, and F. B. V. Waeyenberge, The design and verification of MuMax3, \href{https://doi.org/10.1063/1.4899186}{AIP Adv. \textbf{4}, 107133 (2014)}.

\bibitem{Co} J. Sampaio, V. Cros, S. Rohart, A. Thiaville, and A. Fert, Nucleation, stability and current-induced motion of isolated magnetic skyrmions in nanostructures, \href{https://doi.org/10.1038/nnano.2013.210}{Nat. Nanotech. \textbf{8}, 839 (2013)}.
\bibitem{Venkat2018} G. Venkat, H. Fangohr, and A. Prabhakar, Absorbing boundary layers for spin wave micromagnetics, \href{https://doi.org/10.1016/j.jmmm.2017.06.057}{J. Magn. Magn. Mater. \textbf{450}, 34 (2018)}.

\bibitem{Seb2015} T. Sebastian, K. Schultheiss, B. Obry, B. Hillebrands, and H. Schultheiss, Micro-focused Brillouin light scattering: imaging spin waves at the nanoscale, \href{https://doi.org/10.3389/fphy.2015.00035}{Front. Phys. \textbf{3}, 35 (2015)}.

\bibitem{Kra2018} V. P. Kravchuk, D. D. Sheka, U. K. R\"{o}{\ss}ler, J. van den Brink, and Y. Gaididei, Spin eigenmodes of magnetic skyrmions and the problem of the effective skyrmion mass, \href{https://doi.org/10.1103/PhysRevB.97.064403}{Phys. Rev. B \textbf{97}, 064403 (2018)}.
\bibitem{Muralidhar2021} S. Muralidhar, R. Khymyn, A. A. Awad, A. Alem\'{a}n, D. Hanstorp, and J. {\AA}kerman, Femtosecond Laser Pulse Driven Caustic Spin Wave Beams, \href{https://doi.org/10.1103/PhysRevLett.126.037204}{Phys. Rev. Lett. 126, 037204 (2021)}.
\bibitem{Liu2020} Z.-X. Liu and H. Xiong, Magnon laser based on Brillouin light scattering, \href{https://doi.org/10.1364/OL.401689}{Opt. Lett. \textbf{45}, 5452 (2020)}.
\bibitem{Andrianov2014} S. N. Andrianov and S. A. Moiseev, Magnon qubit and quantum computing on magnon Bose-Einstein condensates, \href{https://doi.org/10.1103/PhysRevA.90.042303}{Phys. Rev. A \textbf{90}, 042303 (2014)}.
\bibitem{Francis2020} A. Francis, J. K. Freericks, and A. F. Kemper, Quantum computation of magnon spectra, \href{https://doi.org/10.1103/PhysRevB.101.014411}{Phys. Rev. B \textbf{101}, 014411 (2020)}.
\bibitem{Han2020} J. Han, P. Zhang, J. T. Hou, S. A. Siddiqui, and L. Liu, Mutual control of coherent spin waves and magnetic domain walls in a magnonic device, \href{https://doi.org/10.1126/science.aau2610}{Science \textbf{366}, 1121 (2019)}.
\bibitem{Hula2021} T. Hula, K. Schultheiss, F. J. T. Goncalves, L. K\"{o}rber, M. Bejarano, M. Copus, L. Flacke, L. Liensberger, A. Buzdakov, A. K\'{a}kay, M. Weiler, R. Camley, J. Fassbender, and H. Schulthei{\ss}, Spin-wave frequency combs, \href{https://arxiv.org/abs/2104.11491v1}{arXiv:2104.11491 (2021)}.
\end{thebibliography}
\end{document}